\documentclass[sts,preprint]{imsart-edit}

\RequirePackage{amsthm,amsmath,amsfonts,amssymb}
\RequirePackage[authoryear]{natbib}
\RequirePackage[colorlinks,citecolor=blue,urlcolor=blue]{hyperref}
\RequirePackage{graphicx}
\RequirePackage{enumerate}
\startlocaldefs

\newcommand{\logit}{\mbox{logit}}

\newcommand{\E}{\mbox{E}}
\newcommand{\Prob}{\mbox{P}}
\newcommand{\V}{\mbox{V}}

\newcommand{\mathspace}[1] {\mathcal{#1}}
\newcommand{\matrixobs}[1] {\mathbf{#1}}
\newcommand{\vectorobs}[1] {\mathrm{#1}}
\newcommand{\scalarobs}[1] {#1}
\newcommand{\RV}[1] {#1}

\newcommand{\R}{\mathbb{R}}
\newcommand{\Xspace}{\mathspace{X}}
\newcommand{\rvX}{\RV{X}}
\newcommand{\rvY}{\RV{Y}}
\newcommand{\rvZ}{\RV{Z}}
\newcommand{\rvM}{\RV{M}}
\newcommand{\rvN}{\RV{N}}
\newcommand{\obsXmat}{\matrixobs{X}}
\newcommand{\obsXvec}{\vectorobs{x}}

\newcommand{\obsZ}{\vectorobs{z}}

\newcommand{\obsN}{\scalarobs{n}}
\endlocaldefs

\begin{document}

\begin{frontmatter}
\title{Semi-supervised learning and the question of true versus estimated propensity scores}
\runtitle{Semi-supervised machine learning for treatment effect estimation}

\begin{aug}
\author[A]{\fnms{Andrew} \snm{Herren}\ead[label=e1]{asherren@asu.edu}}
\and
\author[B]{\fnms{P. Richard} \snm{Hahn}\ead[label=e2]{prhahn@asu.edu}}
\affiliation{Arizona State University}
\end{aug}

\begin{abstract}
A straightforward application of semi-supervised machine learning to the problem of treatment 
effect estimation would be to consider data as ``unlabeled" if treatment assignment and covariates 
are observed but outcomes are unobserved. According to this formulation, large unlabeled 
data sets could be used to estimate a high dimensional propensity function and causal inference 
using a much smaller labeled data set could proceed via weighted estimators using the learned 
propensity scores. In the limiting case of infinite unlabeled data, one may estimate the high 
dimensional propensity function exactly. However, longstanding advice in the causal inference 
community suggests that estimated propensity scores (from labeled data alone) are actually preferable 
to true propensity scores, implying that the unlabeled data is actually useless in this context. In this 
paper we examine this paradox and propose a simple procedure that reconciles the strong intuition 
that a known propensity functions should be useful for estimating treatment effects with the previous 
literature suggesting otherwise. Further, simulation studies suggest that direct regression may be 
preferable to inverse-propensity weight estimators in many circumstances.
\end{abstract}


\end{frontmatter}

\section{Introduction}

In their seminal paper, \cite{rosenbaum1983central} show that the propensity score, the probability of 
receiving the treatment conditional on control covariates, is a sufficient statistic for estimating treatment effects. 
They show that rather than needing to balance on a potentially high dimensional vector of control variables, 
one need only balance on the one-dimensional propensity score. In practice, however, researchers do not 
know the propensity function  --- the mapping from the covariates to the treatment probabilities --- in advance; 
it must estimate from data. Unfortunately, learning the propensity function can be quite difficult when there are 
many controls and its parametric form is unknown. Linear logistic regression is often the uncritical method of 
first resort and even in that familiar setting estimation can be challenging with limited data.  

Notably, propensity score approaches only require learning a function of the treatment assignment and control 
variables; the response variable is not utilized at this preliminary stage. This raises the possibility of incorporating 
side data for which responses aren't measured, so-called ``unlabeled'' data, in the parlance of semi-supervised learning 
(\cite{zhu2009introduction}) to assist in estimating the propensity function. Provided that such data is available, the 
propensity function can be inferred more accurately than would be possible with the labeled data alone. In the limit, one 
can imagine learning the propensity function arbitrarily accurately, thereby realizing the holy grail conveyed by 
\cite{rosenbaum1983central} of being able to do causal inference with a one-dimensional control variable. 

However, it was famously shown by \cite{hirano2003efficient} that true propensity scores are actually worse -- 
in the sense of yielding asymptotically higher variance estimates of the treatment effect -- than ones estimated from 
the labeled data alone! This observation would seem to dash any hope of leveraging unlabeled data as described above. 
Similar results were demonstrated by \cite{lunceford2004stratification} in the context of stratification and weighting 
estimators with parametric propensity models.

This paper unpacks this apparent paradox in an effort to provide practical guidance to applied researchers interested in 
estimating treatment effects from observational data. We find that the uselessness of the true propensity function have 
been greatly exaggerated, reviving the potential for unlabeled data to assist in propensity function estimation, i.e. 
semi-supervised treatment effect estimation.

\section{Problem Statement and Notation}

\subsection{Notation}
\label{subsection:notation}

Throughout the paper we use the following conventions:
\begin{itemize}
\item \textbf{Random variable}: italicized, uppercase letter (e.g. $\RV{X}$, $\RV{A}$, ...)
\item \textbf{Scalar realization of random variable}: italicized, lowercase letter (e.g. $\scalarobs{x}$, $\scalarobs{a}$, ...)
\item \textbf{Vector realization of random variable}: lowercase letter (e.g. $\vectorobs{x}$, $\vectorobs{a}$, ...)
\item \textbf{Matrix realization of random variable}: bold, uppercase letter (e.g. $\matrixobs{X}$, $\matrixobs{A}$, ...)
\item \textbf{Metric space of random variable's support}: math calligraphy letter (e.g. $\mathspace{X}$, $\mathspace{A}$, ...)
\end{itemize}

We first draw a distinction between observational and experimental studies for treatment effect estimation. 
In experimental settings, the treatment of interest is deliberately randomized according to a pre-specified design, 
typically with the goal of enabling straightforward identification of the treatment effect. In observational settings, 
data on treatment, outcome and covariates are observed without any direct manipulation. Researchers wishing 
to infer treatment effects with observational data must rely on a series of assumptions, which we now introduce.

Let $\rvY$ refer to the outcome of interest, $\rvZ$ denote a binary treatment assignment, 
and $\rvX$ be a vector of covariates drawn from covariate space $\Xspace$. 
We use the potential outcomes notation $\rvY^1$ and $\rvY^0$ to refer to the counterfactual values of 
$\rvY$ when $\rvZ=1$ and $\rvZ=0$ (\cite{hernan2020causal}). 
Our estimand of interest, is the average treatment effect: $\E[\rvY^1 - \rvY^0]$.

Since the potential outcomes $(\rvY^1, \rvY^0)$ are never observed simultaneously, we rely on a number of identifying assumptions 
common to the causal inference literature (\cite{rubin1980randomization}, \cite{rosenbaum1983central}, \cite{hernan2020causal}):

\begin{enumerate}
\item Stable unit treatment value assumption (SUTVA): for any sample of size $\obsN$ with $\rvY \in \mathcal{Y}$ 
and $\rvZ \in \mathcal{Z}$, $(\rvY_i^1, \rvY_i^0) \perp \rvZ_j$ for all $i,j \in \{1, ..., \obsN\}$ with $j \neq i$
\item Positivity: $0 < \Prob(\rvZ=1 \mid \rvX=\obsXmat) < 1$ for all $\obsXmat \in \mathcal{X}$
\item Conditional exchangeability: $(\rvY^1, \rvY^0) \perp \rvZ \mid \rvX$
\end{enumerate}

Taken together, these assumptions enable identification of average treatment effects using observational 
data after adjusting for the effect of $\rvX$ on $\rvZ$ and $\rvY$. Thus, we can rewrite our estimand as 
$\textrm{ATE} = \E[\rvY^1 - \rvY^0] = \E_X[\E[\rvY \mid \rvX, \rvZ = 1] - \E[\rvY \mid \rvX, \rvZ = 0]]$. 
There are many ways to adjust for $\rvX$ in estimating the ATE, but this paper focuses on methods that use 
the \textrm{propensity score}.
\cite{rosenbaum1983central} define the propensity score as $p(\rvX) = \Prob(\rvZ = 1 \mid \rvX)$ and show 
that $p(\rvX)$ has several desirable properties:

\begin{enumerate}
\item $\rvX \perp \rvZ \mid p(\rvX)$
\item $(\rvY^1, \rvY^0) \perp \rvZ \mid p(\rvX)$
\end{enumerate}

In short, the propensity score can be used to ensure covariate balance across treatment groups 
and the resulting covariate balance deconfounds the treatment effect estimate of $\rvZ$ on $\rvY$. 
The deconfounding property of $p(\rvX)$ implies that the treatment effect estimand can be rewritten 
as $\textrm{ATE} = \E[\rvY^1 - \rvY^0] = \E_{p(\rvX)}[\E[\rvY \mid p(\rvX), \rvZ = 1] - \E[\rvY \mid p(\rvX), \rvZ = 0]]$.

\subsection{Confounders, Instruments and Pure Prognostic / Moderator Variables}

Above, we let $X$ refer to the set of covariates of $Y$ and $Z$, but for the purposes of this discussion, 
we introduce a taxonomy of covariate types. First, observe that, using a similar framing as 
\cite{hahn2020bayesian}, the relationship between $Y$, $Z$ and $X$ can be expressed as:
\begin{equation*}
\begin{aligned}
\E\left[ Y \mid X_{\mu}, X_{\tau}, X_{\pi} \right] &= \mu\left( X_{\mu} \right) + \tau\left( X_{\tau} \right) Z\\
\Prob\left(Z = 1 \mid X_{\pi} \right) &= \pi \left( X_{\pi} \right)
\end{aligned}
\end{equation*}
where $X_{\mu}$ refers to the set of prognostic variables, $X_{\tau}$ refers to the set of moderator variables, 
and $X_{\pi}$ refers to the set of propensity variables. We consider that any variable in $X$ can be classified 
as follows:

\begin{enumerate}
\item \textbf{Confounder}: Any variable which appears in $X_{\pi}$ and either of $X_{\mu}$ or $X_{\tau}$
\item \textbf{Pure prognostic / moderator variable}: Any variable which appears in either $X_{\mu}$, $X_{\tau}$, or 
both, but not in $X_{\pi}$
\item \textbf{Instrument}: Any variable which appears in $X_{\pi}$ and not in $X_{\mu}$ nor $X_{\tau}$ 
(this is a ``pure propensity'' variable)
\item \textbf{Extraneous variable}: Any variable in $X$ which appears in none of $X_{\tau}$, $X_{\mu}$, or $X_{\pi}$)
\end{enumerate}

\subsection{Estimated vs true propensities} \label{trueprop}

\cite{hirano2003efficient} discuss the inverse-propensity weighted estimator of average treatment effect, 
which estimates the ATE as follows
\[ \textrm{ATE}_{\textrm{IPW}} = \left( \frac{1}{\rvN} \sum_{i=1}^{\rvN} \frac{\rvY_i \rvZ_i}{p(\rvX_i)} - 
\frac{\rvY_i (1-\rvZ_i)}{1-p(\rvX_i)} \right)\] 

They show that even when the true propensities are known, using them directly in the IPW estimator is 
asymptotically inefficient. 
We consider the finite sample variance properties of this estimator on a simple data generating 
process, which we define below:
\begin{itemize}
\item \textbf{Covariates}: $\rvX_1, \rvX_2 \sim \mbox{Bernoulli}(1/2)$, letting $\rvX = (\rvX_1, \rvX_2)$
\item \textbf{Propensity function}: $p(\rvX) = 1 / \left( 1+\exp\{ - ( \alpha_1 + \alpha_2 \rvX_1 ) \} \right)$
\item \textbf{Treatment}: $\rvZ \sim \mbox{Bernoulli}(p(\rvX))$
\item \textbf{Outcome}: $\rvY \sim \mathcal{N}(\gamma_1 \rvX_1 + \gamma_2 \rvX_2 + \tau \rvZ, 1)$
\end{itemize}

We let the random variable $\rvN$ denote the overall sample size and define subset-specific sample sizes as follows:
\begin{itemize}
\item $\rvN_{x}$: the number of observations with $\rvX = x$
\item $\rvN_{x, z}$: the number of observations with $\rvX = x$ and $\rvZ = z$
\end{itemize}

First, observe in this case that $p(\rvX)$ takes on two distinct values: 
$1 / \left( 1 + \exp \{ - \alpha_1 \} \right)$ if $\rvX_1 = 0$, 
and $1 / \left( 1 + \exp \{ - \alpha_1 - \alpha_2 \} \right)$ if $\rvX_1 = 1$.
We can split in the sum into eight parts, stratifying on the unique values of ($\rvX_1$, $\rvX_2$, $\rvZ$). 
For a given stratum, the weights in the 
denominator, as well as the $\rvZ_i$ and $1 - \rvZ_i$ indicators, are all the same and can be factored out. 
We refer to ``true propensities'' estimator as $\mbox{IPW}_{p(\rvX)}$ to indicate that the true propensity 
function, $p(\rvX)$ is the basis for the weights in the denominator.
Let $p_{x} = \Prob(\rvZ = 1 \mid \rvX=x) = \Prob(\rvZ = 1 \mid \rvX_1=x_1)$.
The portion of the ``true propensities'' estimator represented by a given $\{\rvX = x\}$ 
(i.e. $\{\rvX_1 = x_1, \rvX_2 = x_2\}$) stratum is given by
\newpage
\begin{equation*}
\begin{aligned}
\mbox{IPW}_{p(\rvX), x} &= \frac{1}{\rvN} \sum_{j: \rvX_j = x} \left(  \frac{\rvY_j \rvZ_j}{p_x} - 
\frac{\rvY_j (1-\rvZ_j)}{1-p_x} \right)\\ 
&= \frac{1}{\rvN} \frac{\rvN_{x}}{\rvN_{x}} \frac{\rvN_{x, z}}{\rvN_{x, z}} \frac{1}{p_x} \sum_{j: \rvX_j = x, \rvZ_j=1} \rvY_j\\
&\;\;\;\; - \frac{1}{\rvN}\frac{\rvN_{x}}{\rvN_{x}} \frac{\rvN_{x, z=0}}{\rvN_{x, z=0}} \frac{(1-p_x)}{(1-p_x)} \sum_{j: \rvX_j = x, \rvZ_j=0} \rvY_j\\
&= \left( \frac{\rvN_x}{\rvN} \right) \left( \frac{\hat{p}_x}{p_x} \bar{\rvY}_{x, \rvZ=1} - 
\frac{1-\hat{p}_x}{1-p_x} \bar{\rvY}_{x, \rvZ=0} \right)
\end{aligned}
\end{equation*}
Where $\mbox{IPW}_{p(\rvX)} = \mbox{IPW}_{p(\rvX), \rvX=(0, 0)} + \mbox{IPW}_{p(\rvX), \rvX=(1, 0)} 
+ \mbox{IPW}_{p(\rvX), \rvX=(0, 1)} + \mbox{IPW}_{p(\rvX), \rvX=(1, 1)}$ 
and $\hat{p}_x = \left( \rvN_{x, \obsZ=1} / \rvN_{x} \right)$. 
We see that replacing the true propensity weights with these empirical weights results in the following estimator:
\begin{equation*}
\begin{aligned}
\mbox{IPW}_{p(\rvX), x} &= \frac{1}{\rvN} \sum_{j: \rvX_j = x} \left(  \frac{\rvY_j \rvZ_j}{\hat{p}_x} - 
\frac{\rvY_j (1-\rvZ_j)}{1-\hat{p}_x} \right)\\ 
&= \left( \frac{\rvN_x}{\rvN} \right) \left( \bar{\rvY}_{x, \rvZ=1} - \bar{\rvY}_{x, \rvZ=0} \right)
\end{aligned}
\end{equation*}

We can compare the variances of these two estimators, by first observing that the data are i.i.d., so that 
conditional on the number of observations with $\rvZ=1$ and $\rvX = x$, the covariances of the strata means are 0.
We derive the variances of the two estimators in Appendix \ref{appA}, but the key to understanding why 
$\V\left[ \mbox{IPW}_{\hat{p}(\rvX), x} \right] \leq \V\left[ \mbox{IPW}_{p(\rvX), x} \right]$ is to notice that the ratios 
$\left( \hat{p}_{x_1} / p_{x_1} \right)$ and $\left( 1-\hat{p}_x \right) / \left(1-p_x \right)$ are random variables 
which converge to $1$ as $n \rightarrow \infty$ but are noisy in finite samples. 
These finite sample imbalances are an ancillary statistic for $\tau$, 
and the $\hat{p}(\rvX)$ estimator conditions on this ancillary statistic for more efficient inference.
For a detailed discussion of the role of ancillary statistics in causal inference, readers are referred to 
Chapter 10 of \cite{hernan2020causal} and the references therein. 

This result seems to imply a paradox. Even if a propensity function is known exactly, is the analyst better off 
ignoring that information and estimating the propensity scores? While it is true that directly weighting by the 
true propensities yields higher variances, we show that the true propensity scores can still be used to achieve 
efficiency. Consider in this case that $p(\rvX) = 1 / \left( 1 + \exp \{ - \alpha_1 - \alpha_2 \rvX_1 \} \right)$, so 
$\rvZ$ can be modeled via logistic regression on $\rvX_1$, $\rvX_2$, and an intercept term. 
Fitting this model on a sample $\obsXmat = (\obsXvec_1, \obsXvec_2)$ and $\obsZ$ would yield coefficient estimates 
$\hat{\beta}_1$, $\hat{\beta}_2,$ and $\hat{\beta}_3$ such that 
$\hat{p}(\obsXmat) = \frac{1}{1+\exp\{- \hat{\beta}_1 - \hat{\beta}_2 \obsXvec_1 - \hat{\beta}_3 \obsXvec_2\}}$, where the true values of 
the parameters are $\beta_1 = \alpha_1$, $\beta_2 = \alpha_2$, and $\beta_3 = 0$. 
Thus, if we know the true value of $p(\obsXmat)$, 
we can calculate the true $\beta_1 + \beta_2 \obsXvec_1 + \beta_3 \obsXvec_2$ by logit transformation 
\[\beta_1 + \beta_2 \obsXvec_1 + \beta_3 \obsXvec_2 = -\log\left( \frac{1}{p(\obsXmat)} - 1 \right)\]
Since both $\beta_1 + \beta_2 \obsXvec_1 + \beta_3 \obsXvec_2$ and 
$\hat{\beta}_1 + \hat{\beta}_2 \obsXvec_1 + \hat{\beta}_3 \obsXvec_2$ are one-dimensional linear functions of $\obsXmat$, 
we could express $\hat{\beta}_1 - \hat{\beta}_2 \obsXvec_1 - \hat{\beta}_3 \obsXvec_2 = \hat{\eta}_0 
 - \log\left( \frac{1}{p(\obsXmat)} - 1 \right) \hat{\eta}_1$ for some 
$\hat{\eta}_0, \hat{\eta}_1 \in \R$ and observe that this is equivalent to a logistic regression of $Z$ on a 1-dimensional 
vector $-\log\left( \frac{1}{p(\obsXmat)} - 1 \right)$ with an intercept term.

This 1-dimensional regression adjustment is similar to the sample correction discussed in the context of the $\hat{p}(\rvX)$ 
estimator, with one key difference: the differences between true and empirical treatment probabilities are adjusted across the 
support of $p(\rvX)$ rather than $\rvX$. 
In our two-covariate example, the difference is that $p(\rvX)$ is only a function of $\rvX_1$, so has two unique values, 
rather than the four distinct values of $\rvX$.

We introduce the following shorthand to refer to the three estimators:
\begin{enumerate}
\item $p(\rvX)$: IPW estimator using the true propensity function, $p(\rvX)$ as weights
\item $\hat{p}(\rvX)$: IPW estimator using the covariate-estimated propensity function, 
$\hat{p}(\rvX)$, as weights
\item $\hat{p}(p(\rvX))$: IPW estimator using the true propensity function, $p(\rvX)$, as a 1-dimensional variable for 
estimating sample propensity weights
\end{enumerate}

\cite{hirano2003efficient} and the example above demonstrate that the $\hat{p}(\rvX)$ estimator is lower variance than the 
$p(\rvX)$ estimator. However, we see in Appendix \ref{appB} that the $\hat{p}(p(\rvX))$ estimator has lower variance than 
the $\hat{p}(\rvX)$ estimator as long as the following two conditions are satisfied:
\begin{itemize}
\item $\rvX_2$ is not a pure prognostic / moderator variable: 
$\E \left[\bar{\rvY}_{\rvX_1 = x_1, \rvX_2 = 1, \rvZ=z}\right] = \E \left[\bar{\rvY}_{\rvX_1 = x_1, \rvX_2 = 0, \rvZ=z}\right]$ 
for all $x_1, z$. In our simple data generating process, this is true if $\gamma_2 = 0$.
\item The outcome variance does not differ along the support of $\rvX_2$: 
$\V \left[\bar{\rvY}_{\rvX_1 = x_1, \rvX_2 = 1, \rvZ=z}\right] = \V \left[\bar{\rvY}_{\rvX_1 = x_1, \rvX_2 = 0, \rvZ=z}\right]$ 
for all $x_1, z$. In our simple data generating process, this is true as the variance of $Y$ is constant with respect to $X$.
\end{itemize}
The most interesting takeaway from this result is that the cases in which the $\hat{p}(\rvX)$ 
estimator outperforms the $\hat{p}(p(\rvX))$ estimator do not pertain to the intended use of the 
propensity score, to deconfound treatment assignment and enable identification of the ATE. 
The $\hat{p}(\rvX)$ estimator can attain a lower variance, but it does so by stratifying on a pure
prognostic / moderator variable, $\rvX_2$. This points to an unintended use of weighting / stratification 
estimators, rather than a fundamental problem with using true propensities. Adjusting for pure prognostic 
/ moderator variables can be achieved by direct regression adjustment, and \cite{hahn2020bayesian} 
provide a nonparametric Bayesian approach that shows promising results on a number of difficult data 
generating processes.

To demonstrate the difference in variances of the three estimators in finite samples, we conduct a simulation study.
We use the same data generating process discussed in the finite sample variance calculations above, in which 
$\rvY$ is normal, $\rvX$ and $\rvZ$ are Bernoulli, and $\rvX$ has $p = 2$ dimensions.
In the simulations presented below, we generate 10,000 datasets for each value of $\rvN$ 
and estimate the average treatment effect using the IPW estimator, comparing three possible weighting options:

\begin{enumerate}[a)]
\item $p(\rvX)$: True propensities ($\{0.3, 0.7\}$)
\item $\hat{p}(\rvX)$: Estimated propensities (logistic regression of $\rvZ$ on $\rvX$)
\item $\hat{p}(p(\rvX))$: Sample adjustment using true propensities 
(logistic regression of $\rvZ$ on $\logit\left(p(\rvX)\right)$)
\end{enumerate}

\begin{table}[ht]
\centering
\begin{tabular}{rrrr}
  \hline
$\rvN$ & $p(\rvX)$ & $\hat{p}(\rvX)$ & $\hat{p}(p(\rvX))$ \\ 
  \hline
15 & 1.80 & 0.91 & 0.81 \\ 
25 & 1.41 & 0.69 & 0.52 \\ 
50 & 1.00 & 0.38 & 0.32 \\ 
100 & 0.70 & 0.24 & 0.22 \\ 
250 & 0.44 & 0.14 & 0.14 \\ 
   \hline
\end{tabular}
\caption{Standard deviation of IPW estimates using different weighting approaches\protect\\
10,000 simulations per $\rvN$\protect\\
p = 2 (value of $\rvN$ for which variances converge depends on p)} 
\end{table}

Our simulations confirm the points made in \cite{hirano2003efficient} and \cite{rosenbaum1987model} 
that using the true propensities directly leads to higher variances in finite samples. 
However, we also see that the true propensities can be used in a one-dimensional model of $\rvZ$ to attain 
even lower variances than estimators using the fitted propensities. 

\subsection{Semi-supervised treatment effect estimation}

In practice, knowledge of the exact true propensity function is unlikely outside of randomized experiments.
However, it is certainly possible to imagine investigators having access to auxiliary data that enable more 
accurate estimates of the true propensities. The received wisdom of \cite{hirano2003efficient}, 
\cite{lunceford2004stratification}, and \cite{rosenbaum1987model} would seem to imply that such accuracy 
gains impede efficient estimation of the ATE. The primary challenge in using propensity scores estimated 
from a larger auxiliary dataset stems from the utility of conditioning on finite sample treatment probabilities in weighting 
estimators. However, we have shown in Section \ref{trueprop} that the same variance reduction can be 
achieved by conditioning on finite sample probabilities formed by $p(\rvX)$ rather than $\rvX$, and this 
result extends to an estimate of $p(\rvX)$ derived from arbitrarily large datasets.

We can formalize the notion of using a large auxiliary dataset in estimating the ATE by framing it as a 
semi-supervised learning problem. 
Semi-supervised learning refers to a broad class of techniques that combine labeled and unlabeled data 
in statistical learning problems (\cite{zhu2009introduction}, \cite{belkin2006manifold}, \cite{liang2007use}). 
Consider a dataset with $\rvN$ observations. As above, we let $\rvY$, $\rvZ$, and $\rvX$ denote outcome, 
treatment, and covariates, respectively. We refer to the data points with observed outcomes as ``labeled data," 
and let $\rvN_o$ refer to the number of such labeled observations and $\rvN_m = \rvN - \rvN_o$ refer 
to the number of data points with missing outcomes.
We index an outcome's status as missing or observed by $\rvM$.

Following the notation of \cite{liang2007use}, we denote ``unlabeled,'' or marginal, data as
\begin{itemize}
\item $\rvX^m = \{\rvX_i; i = \rvN_o + 1, ... , \rvN_o + \rvN_m \}$, and 
\item $\rvZ^m = \{\rvZ_i; i = \rvN_o + 1, ... , \rvN_o + \rvN_m \}$
\end{itemize}
Similarly, we denote ``labeled'' data as
\begin{itemize}
\item $\rvY^o = \{\rvY_i; i = 1, ... , \rvN_o \}$, 
\item $\rvX^o = \{\rvX_i; i = 1, ... , \rvN_o \}$, and 
\item $\rvZ^o = \{\rvZ_i; i = 1, ... , \rvN_o \}$,
\end{itemize}
We observe that $\rvM = 1$ where $\rvX, \rvZ \in (\rvX^m, \rvZ^m)$ and $\rvM = 0$ elsewhere.
The discussion in Section \ref{trueprop} shows that the set of marginal $(\rvX^m, \rvZ^m)$ data can be put to 
good use in estimating the average treatment effect as follows:

\begin{enumerate}
\item Obtain an estimate, $p_{e}(\rvX)$ of the true propensities using the labeled and unlabeled data $(\rvX, \rvZ)$
\item Adjust the estimated propensities to the labeled sample by modeling 
$\rvZ^o \sim \mbox{Bernoulli} \left( p_{e}(\rvX^o) \right)$, and denote these adjusted estimates by 
$p^{*}(\rvX^o)$
\item Estimate the ATE using only the labeled data ($\rvY^o$, $\rvZ^o$, and $p^{*}(\rvX^o)$)
\end{enumerate}

As with other constructions in the semi-supervised learning literature, this framing allows us to profitably use information about the statistical patterns in the unlabeled data 
(in this case, $p(\rvZ = 1 \mid \rvX)$) to better estimate an effect among the labeled data ($\E[\rvY^1 - \rvY^0]$).

\section{Simulations}
\label{section:simstudy}

We conduct several simulation studies to understand the phenomena discussed above in more depth. 
While logistic regression is often a natural choice for estimating propensity scores, \cite{lee2010improving} 
point out that the assumptions required for logistic regression are not always warranted and they examine 
propensity score estimation using a number of machine learning methods, including CART 
(\cite{breiman1984classification}) and Random Forest (\cite{breiman2001random}).

This paper proceeds similarly, but instead relies on Bayesian Additive Regression Trees (BART) 
(\cite{chipman2010bart}).
\footnote{Code can be found at https://github.com/andrewherren/semi-supervised-propensity} 
BART is a nonparametric Bayesian tree ensemble method that estimates complex functions via a 
sum of weak regression or classification trees. In the case of our IPW estimator, we also consider 
a logistic propensity model in order to investigate how inferences can be harmed by model mis-specification.

\subsection{Estimators}

We employ three estimators which use propensity scores in our simulation study:

\begin{itemize}
\item Inverse Propensity Weighted (IPW) estimator (\cite{hirano2003efficient}), introduced in Section \ref{trueprop} and defined as
\[ \textrm{ATE}_{\textrm{IPW}} = \frac{1}{\rvN} \sum_{i = 1}^{\rvN} \left( \frac{\rvY_i \rvZ_i}{p(\rvX_i)} - 
\frac{\rvY_i (1 - \rvZ_i)}{(1 - p(\rvX_i))} \right) \]

\item Targeted Maximum Likelihood Estimator (TMLE) (\cite{van2010targeted}, \cite{van2010targeted2}, 
\cite{gruber2009targeted})
\[\textrm{ATE}_{\textrm{TMLE}} = \frac{1}{\rvN} \sum_{i = 1}^{\rvN} \left( Q^*\left( \rvZ_i = 1, \rvX_i \right) 
- Q^*\left( \rvZ_i = 0, \rvX_i \right) \right)\]
where $Q^*\left( \rvZ_i, \rvX_i \right)$ represents a semi-parametric model of the outcome $\rvY$ 
which incorporates the propensity score.

\item Bayesian Causal Forests (BCF) (\cite{hahn2020bayesian})
\[\textrm{ATE}_{\textrm{BCF}} = \frac{1}{\rvN} \sum_{i = 1}^{\rvN} \left( \bar{f}(\rvX_i, \rvZ_i = 1) - 
\bar{f}(\rvX_i, \rvZ_i = 0) \right) \]
where $\bar{f}(\rvX_i, \rvZ_i)$ is an average of posterior simulations (across all simulations) of two 
BART models, defined as $f(\rvX_i, \rvZ_i) = \mu(\rvX_i, p(\rvX_i)) + 
\tau(\rvX_i) \rvZ_i$. 
\end{itemize}

\subsection{Simulation approaches}

Given the two step nature of each of the above estimators, in which a propensity model is first constructed 
and its estimates then used to calculate an average treatment effect, we consider three approaches to 
treatment effect estimation:

\begin{itemize}
\item \textbf{Complete case analysis}: We discard all unlabeled data samples, estimate $\hat{p}(\obsXmat^o)$, 
and then compute average treatment effects on only the labeled observations.
\item \textbf{Semi-supervised}: We use the labeled and unlabeled $\obsXmat$ and $\obsZ$ values to 
estimate $\hat{p}(\obsXmat)$ and then use the $\hat{p}(\obsXmat^o)$ predictions to compute average 
treatment effects on the labeled observations.
\item \textbf{True propensities}: Simulation studies provide us with actual probabilities of receiving treatment, 
so we can use these true $p(\obsXmat^o)$ to compute average treatment effects on the labeled observations. 
Of course, in most real world applications, the exact true propensities will not be available, and we 
consider this scenario as a limiting case in which an arbitrarily large amount of unlabeled data were available 
for modeling $\hat{p}(\obsXmat)$. 
\end{itemize}

\subsection{Interval construction}

We construct confidence intervals for our simulations as follows:

\begin{itemize}
	\item \textbf{IPW (logit)} and \textbf{IPW (BART)}: We compute the asymptotic standard error 
	assuming a logistic propensity model as in \cite{cerulli2014treatrew}
	\item \textbf{TMLE}: We use the 95\% confidence interval produced by the \texttt{tmle} R package
	\item \textbf{BCF}: We construct a 95\% credible interval using posterior samples of the treatment effect 
\end{itemize}

\subsection{Evaluation metrics}

We evaluate the results of our simulations using the following metrics.

\begin{itemize}
	\item \textbf{RMSE}: root mean squared error of estimated treatment effects
	\item \textbf{Bias}: difference between true (simulated) ATE and average estimated ATE
	\item \textbf{Coverage}: fraction of 95\% confidence / credible intervals that contain the true ATE
\end{itemize}

\subsection{Data Generating Process Simulations}

For simulated covariates, outcomes, and treatment effects, we use a subset of the data generating 
processes tested in \cite{hahn2020bayesian}. 
\begin{align*}
\rvY &= \mu(\rvX) + \tau \rvZ + \varepsilon\\
\mu(\rvX) &= 
\begin{cases}
3 + \rvX_1 \rvX_3,&  \rvX_5 = 1,\\
\rvX_1 \rvX_3,&  \rvX_5 = 2\\
-3 + x_1 \rvX_3,&  \rvX_5 = 3
\end{cases} \\
\rvX_1, \rvX_2, \rvX_3 &\sim \mathcal{N}(0, 1)\\
\rvX_4 &\sim \textrm{Bernoulli}(0.5)\\
\rvX_5 &\sim \textrm{Categorical}(0.25, 0.5, 0.25)\\
\tau &= 3\\
\varepsilon &\sim \mathcal{N}(0, 1)
\end{align*}
There are $p = 5$ covariates, the first three of which are independent standard normal variables, 
the fourth of which is binary, and the fifth is an unordered categorical variable with values $1, 2, 3$. 
Treatment effects are homogeneous ($\tau = 3$), and the outcome is determined by a combination of 
treatment effects and a piecewise interaction function of three of the covariates
($\mu(\rvX) = 1 + g(\rvX_5) + \rvX_1 \rvX_3$, where $g(\rvX) = 2$ if $\rvX = 1$, $g(\rvX) = -1$ 
if $\rvX = 2$, and $g(\rvX) = -4$ if $x = 3$). 
$\mu(\rvX)$ is often referred to as a \textit{prognostic effect} and can be conceptualized as the expected value of 
the outcome for individuals who do not receive the treatment. 

For treatment assignment, we consider two cases:

\begin{enumerate}
\item $P(\rvZ = 1 \mid \rvX) = 0.5$
\item $P(\rvZ = 1 \mid \rvX) = 0.8\Phi(3 \mu(\rvX) / s - 0.5 \rvX_1) + 0.05 + u / 10$
\begin{itemize}
	\item $\Phi(\cdot)$ is the standard normal CDF
	\item $s$ is the sample standard deviation of simulated observations of $\mu(\rvX)$
	\item $u \sim \mathrm{Uniform}(0, 1)$ 
\end{itemize}
\end{enumerate}

The first treatment assignment mechanism mirrors a simple balanced randomized experiment. 
This presents relatively straightforward inference, as there is no confounding, but it allows us to 
simulate the phenomenon of including non-confounders in the propensity model.
The second treatment assignment mechanism is described as ``Targeted Selection" in 
\cite{hahn2020bayesian}, and refers to the phenomenon of assigning treatment based on the 
expected value of the outcome for those who don't receive treatment. 

In order to simulate the process by which outcomes are unlabeled, we assume a fixed overall sample size 
of 5,000 and randomly label 50, 100, or 500 observations so that $\rvM$ is not correlated with $\rvY$, $\rvX$, or $\rvZ$. 
This parallels the notion of ``missing completely at random'' (MCAR) in the missing data literature 
(\cite{little2019statistical}), though our problem is slightly different. We treat our unlabeled $\rvX^m$ and $\rvZ^m$ 
as auxiliary data that may be helpful in estimating the ATE rather than treating our unobserved $\rvY^m$ as 
missing data to be imputed or otherwise estimated. Before proceeding to a more detailed discussion of the 
simulation results, we briefly summarize the key takeaways:
\begin{itemize}
	\item Using unlabeled data will harm inferences in the case of a mis-specified propensity model, 
	as confidence intervals narrow around a biased estimate
	\item BCF and IPW/BART work best in the case of targeted selection
	\item While randomized treatment assignment makes ATE estimation much easier, 
	unlabeled data can still be useful in this case for variance reduction purposes
\end{itemize}

\subsection{Simulation results under targeted selection}

We first examine the targeted selection case, in which treatment assignment is correlated with the 
prognostic effect. 

\subsubsection{IPW (logit)\\}

The table below shows the results for 500 simulations using an IPW estimator and propensities 
estimated by logistic regression. The targeted selection function is nonlinear and thus 
a logistic propensity model is misspecified. We see that the results are badly biased, 
except those which use the true propensities and are not impacted by the model misspecification. 
Furthermore, we observe that using unlabeled data and increasing sample size is harmful, as the 
variance is reduced around a biased estimate, leading to worse interval coverage.

% latex table generated in R 3.6.2 by xtable 1.8-4 package
% Sat Apr  4 06:53:31 2020
\begin{table}[ht]
\centering
\begingroup\small
\begin{tabular}{lllcccc}
  \hline
Approach & N & \# labeled & ATE & RMSE & Bias & Coverage \\ 
  \hline
  Complete Case & 5,000 &  50 & 3.49 & 0.89 & 0.49 & 0.67 \\ 
  Complete Case & 5,000 & 100 & 3.48 & 0.68 & 0.48 & 0.77 \\ 
  Complete Case & 5,000 & 500 & 3.50 & 0.51 & 0.50 & 0.59 \\ 
  Semi-supervised & 5,000 &  50 & 3.48 & 0.97 & 0.48 & 0.67 \\ 
  Semi-supervised & 5,000 & 100 & 3.56 & 0.73 & 0.56 & 0.72 \\ 
  Semi-supervised & 5,000 & 500 & 3.50 & 0.52 & 0.50 & 0.55 \\ 
  True propensities & 5,000 & 500 & 2.99 & 0.25 & -0.01 & 0.89 \\ 
   \hline
\end{tabular}
\endgroup
\caption{500 simulations with outcomes MCAR} 
\end{table}

\subsubsection{IPW (BART)\\}

The table below shows the results for 500 simulations using an IPW estimator and propensities 
estimated by BART. In this case, the propensity model is not mis-specified, and we observe that 
the use of unlabeled data reduces bias and increases interval coverage.

% latex table generated in R 3.6.2 by xtable 1.8-4 package
% Sat Apr  4 06:53:31 2020
\begin{table}[ht]
\centering
\begingroup\small
\begin{tabular}{lllcccc}
  \hline
Approach & N & \# labeled & ATE & RMSE & Bias & Coverage \\ 
  \hline
  Complete Case & 5,000 &  50 & 3.90 & 0.90 & 0.90 & 0.47 \\ 
  Complete Case & 5,000 & 100 & 3.79 & 0.79 & 0.79 & 0.39 \\ 
Complete Case & 5,000 & 500 & 3.50 & 0.50 & 0.50 & 0.36 \\ 
  Semi-supervised & 5,000 &  50 & 3.00 & 0.85 & -0.00 & 0.73 \\ 
  Semi-supervised & 5,000 & 100 & 3.00 & 0.58 & -0.00 & 0.86 \\ 
  Semi-supervised & 5,000 & 500 & 3.00 & 0.26 & -0.00 & 0.90 \\ 
  True propensities & 5,000 & 500 & 2.99 & 0.25 & -0.01 & 0.89 \\ 
   \hline
\end{tabular}
\endgroup
\caption{500 simulations with outcomes MCAR} 
\end{table}

\newpage
\subsubsection{TMLE\\}

The table below shows the results for 500 simulations using a TMLE estimator and propensities 
estimated by BART. In this case, the propensity model is not mis-specified, however, we observe 
that the TMLE estimator achieves poor interval coverage, even in large sample sizes where the 
average bias is lower.

% latex table generated in R 3.6.2 by xtable 1.8-4 package
% Sat Apr  4 06:53:31 2020
\begin{table}[ht]
\centering
\begingroup\small
\begin{tabular}{lllcccc}
  \hline
Approach & N & \# labeled & ATE & RMSE & Bias & Coverage \\ 
  \hline
  Complete Case & 5,000 &  50 & 3.58 & 0.62 & 0.58 & 0.35 \\ 
  Complete Case & 5,000 & 100 & 3.44 & 0.47 & 0.44 & 0.33 \\ 
  Complete Case & 5,000 & 500 & 3.12 & 0.15 & 0.12 & 0.63 \\ 
  Semi-supervised & 5,000 &  50 & 3.33 & 0.51 & 0.33 & 0.60 \\ 
  Semi-supervised & 5,000 & 100 & 3.24 & 0.35 & 0.24 & 0.64 \\ 
  Semi-supervised & 5,000 & 500 & 3.07 & 0.13 & 0.07 & 0.80 \\ 
  True propensities & 5,000 & 500 & 3.05 & 0.12 & 0.05 & 0.84 \\ 
   \hline
\end{tabular}
\endgroup
\caption{500 simulations with outcomes MCAR} 
\end{table}

\subsubsection{BCF\\}

The table below shows the results for 500 simulations using a BCF estimator and propensities 
estimated by BART. BCF was designed in part to handle the case of targeted selection, and we 
see that it produces unbiased estimates with high coverage as more unlabeled data is incorporated.

% latex table generated in R 3.6.2 by xtable 1.8-4 package
% Sat Apr  4 06:53:31 2020
\begin{table}[ht]
\centering
\begingroup\small
\begin{tabular}{lllcccc}
  \hline
Approach & N & \# labeled & ATE & RMSE & Bias & Coverage \\ 
  \hline
  Complete Case & 5,000 &  50 & 3.54 & 0.99 & 0.54 & 0.70 \\ 
  Complete Case & 5,000 & 100 & 3.43 & 0.80 & 0.43 & 0.80 \\ 
Complete Case & 5,000 & 500 & 3.06 & 0.37 & 0.06 & 0.96 \\ 
  Semi-supervised & 5,000 &  50 & 3.26 & 0.80 & 0.26 & 0.77 \\ 
  Semi-supervised & 5,000 & 100 & 3.15 & 0.58 & 0.15 & 0.86 \\ 
  Semi-supervised & 5,000 & 500 & 3.03 & 0.37 & 0.03 & 0.98 \\ 
  True propensities & 5,000 & 500 & 3.01 & 0.32 & 0.01 & 0.98 \\ 
   \hline
\end{tabular}
\endgroup
\caption{500 simulations with outcomes MCAR} 
\end{table}

\newpage
\subsection{Simulation results under randomized assignment}

We now turn to the randomized treatment assignment scenario. 

\subsubsection{Complete case\\}
The table below shows the results for 500 simulations using the complete case approach 
across all four estimators. We see in this case that all methods produce unbiased estimates 
of the ATE as sample size increases, and with the exception of the TMLE estimator, the approaches 
all achieve high coverage of the true treatment effect.

% latex table generated in R 3.6.2 by xtable 1.8-4 package
% Sat Apr  4 06:53:31 2020
\begin{table}[ht]
\centering
\begingroup\small
\begin{tabular}{lllcccc}
  \hline
Estimator & N & \# labeled & ATE & RMSE & Bias & Coverage \\ 
  \hline
  IPW (logistic) & 5,000 &  50 & 3.03 & 0.39 & 0.03 & 0.99 \\ 
  IPW (logistic) & 5,000 & 100 & 2.97 & 0.23 & -0.03 & 1.00 \\ 
IPW (logistic) & 5,000 & 500 & 3.00 & 0.10 & 0.00 & 1.00 \\ 
  IPW (BART) & 5,000 &  50 & 2.67 & 0.46 & -0.33 & 0.98 \\ 
  IPW (BART) & 5,000 & 100 & 2.75 & 0.31 & -0.25 & 0.99 \\ 
  IPW (BART) & 5,000 & 500 & 2.90 & 0.13 & -0.10 & 1.00 \\ 
  TMLE & 5,000 &  50 & 2.96 & 0.33 & -0.04 & 0.72 \\ 
  TMLE & 5,000 & 100 & 2.96 & 0.20 & -0.04 & 0.79 \\ 
  TMLE & 5,000 & 500 & 3.00 & 0.08 & 0.00 & 0.86 \\ 
  BCF & 5,000 &  50 & 2.90 & 0.69 & -0.10 & 0.79 \\ 
  BCF & 5,000 & 100 & 2.93 & 0.47 & -0.07 & 0.87 \\ 
  BCF & 5,000 & 500 & 2.99 & 0.22 & -0.01 & 0.96 \\ 
   \hline
\end{tabular}
\endgroup
\caption{500 simulations with outcomes MCAR} 
\end{table}

\subsubsection{Semi-supervised\\}
Below are the same simulations using the semi-supervised approach. 
For both of the weighting estimators, we observe that the semi-supervised approach yields a 
higher variance. This is an example of the phenomenon discussed in Section \ref{trueprop} -- 
the sample imbalances among the labeled data are an ancillary statistic for $\tau$ and conditioning 
on that statistic leads to lower variance. We note that this problem could be overcome using a 
sample adjustment of $\rvZ$ on the unlabeled propensity score estimates. We also observe 
in this case that bias of the IPW estimators decreases with the use the unlabeled data, which suggests the 
use of the marginal data for two reasons:
\begin{itemize}
\item Using only labeled data can lead to biased treatment effect estimates in finite samples
\item With an appropriate adjustment, the variance of estimators that use marginal propensities can be 
reduced to mirror estimators using only labeled data
\end{itemize}

\newpage
% latex table generated in R 3.6.2 by xtable 1.8-4 package
% Sat Apr  4 06:53:31 2020
\begin{table}[ht]
\centering
\begingroup\small
\begin{tabular}{lllcccc}
  \hline
Estimator & N & \# labeled & ATE & RMSE & Bias & Coverage \\ 
  \hline
  IPW (logistic) & 5,000 &  50 & 3.00 & 0.65 & -0.00 & 0.90 \\ 
  IPW (logistic) & 5,000 & 100 & 3.00 & 0.47 & -0.00 & 0.93 \\ 
  IPW (logistic) & 5,000 & 500 & 3.00 & 0.20 & 0.00 & 0.97 \\ 
  IPW (BART) & 5,000 &  50 & 2.97 & 0.65 & -0.03 & 0.91 \\ 
  IPW (BART) & 5,000 & 100 & 2.97 & 0.47 & -0.03 & 0.93 \\ 
  IPW (BART) & 5,000 & 500 & 2.98 & 0.20 & -0.02 & 0.96 \\ 
  TMLE & 5,000 &  50 & 2.95 & 0.33 & -0.05 & 0.76 \\ 
  TMLE & 5,000 & 100 & 2.96 & 0.20 & -0.04 & 0.81 \\ 
  TMLE & 5,000 & 500 & 3.00 & 0.08 & 0.00 & 0.87 \\ 
  BCF & 5,000 &  50 & 2.91 & 0.70 & -0.09 & 0.82 \\ 
  BCF & 5,000 & 100 & 2.94 & 0.46 & -0.06 & 0.90 \\ 
  BCF & 5,000 & 500 & 2.99 & 0.22 & -0.01 & 0.97 \\ 
   \hline
\end{tabular}
\endgroup
\caption{500 simulations with outcomes MCAR} 
\end{table}

\section{Discussion}

\subsection{Resolving the estimated propensity paradox}
\label{subsection:resolvingparadox}

Section \ref{trueprop} reviews the paradox of using estimated propensities to reduce variance in weighting estimators, 
even if true propensities are known (cited in \cite{hirano2003efficient}, \cite{rosenbaum1987model}, and 
\cite{lunceford2004stratification}). We believe this issue deserves a detailed discussion, as it contains a 
number of subtleties and its central claim (i.e. that estimated propensities are ``better'' than true propensities from a 
variance perspective) persists in the causal inference community to this day.

\subsubsection{What the paradox actually implies\\}

The mechanism by which conditioning on estimated propensities can reduce variances has two components:
\begin{enumerate}[a)]
\item Adjusting for finite sample differences between true propensities and empirical treatment probabilities
\item (For weighting estimators) effecting a variant of direct regression adjustment by stratifying 
on pure prognostic / moderator variables
\end{enumerate} 

We noted in Section \ref{trueprop} that the variance implications of (a) can be mitigated 
by modeling $\rvZ$ directly from the true propensity score, if it is available.
This leaves (b) as a concern, which is no longer a paradox of which propensity scores to use, but 
rather which estimator to use. In particular, this is not a concern for direct regression estimators which 
allow for adjustment by prognostic / moderator variables. Indeed, our simulations using BCF demonstrate this point.

\subsubsection{Bias-variance tradeoff\\}

Absent from most of the literature on estimated vs true propensities is the possibility of model misspecification 
in estimating propensities. Thus, points (a) and (b) above represent perhaps an optimistic view which is that
the propensity function can be estimated arbitrarily well on finite samples. In this case, concerns about 
variance outweigh concerns about bias. However, our simulations show that in plausible real-world scenarios, 
propensity-based ATE estimators that rely only on labeled data can be severely biased. Indeed, while Section 
\ref{trueprop} shows that the variance of the ``true propensities'' estimator can be higher than either of the 
two adjusted estimators, we see in our simulations that the true propensities estimator typically achieves the 
lowest RMSE, implying that bias overwhelms variance in many cases.

\subsection{Propensity model specification}
\label{subsection:overfitprop}

Though there may be benefits to using estimated propensity scores, in practice, calculating propensity 
scores introduces a new set of risks. The simulations in Section \ref{section:simstudy} show that a 
misspecified propensity model can bias estimates, sometimes drastically. 
It may seem natural, then, to treat a propensity model as any other supervised learning 
problem, with variable selection, model validation, and diagnostics to ensure a proper fit. 
Indeed, \cite{mccaffrey2004propensity} use boosting to estimate propensities, 
in part for the automatic variable selection and flexible model specification that tree ensembles provide. 
Their results show that using boosting for a propensity model leads to better covariate balance and 
lower standard errors than a logistic propensity model.

However, \cite{hernan2020causal} caution that traditional model selection techniques may be of dubious 
utility in constructing propensity models. They note that the purpose of the propensity score is to control 
for confounding variables, not predict $Z$ from $X$ arbitrarily well. 
They are careful to point out that a purely data-driven approach to model selection runs the risk of 
including variables, such as colliders, mediators, or post-treatment variables, that jeopardize identification 
of treatment effects. They also advise that, even when treatment effects are identified, including 
instruments in a propensity model can increase the variance of treatment effect estimates 
by pushing estimated propensities closer to 0 and 1.

At first glance, this point would seem to argue against our recommendation to use semi-supervised 
machine learning in ATE estimation. Indeed, naively overfitting a propensity model is a surefire way 
to get noisy (and perhaps even biased) treatment effect estimates. However, we believe that 
a number of practical steps can be taken to mitigate these concerns. 

First, estimated propensity scores should always be visualized or otherwise inspected to ensure that 
they are not numerically close to 0 and 1. 
Second, regularization methods (such as $\ell^2$ normalization or certain types of ensembles), 
can be used instead of logistic regression to minimize the risk of estimating propensity scores close to 0 and 1. 
Finally, subject matter expertise should always play a role in building propensity models. 
In many applied problems, especially those with a high-dimensional covariate space, 
the challenge of constructing an accurate causal graph is highly non-trivial. Insofar as subject matter 
expertise can identify variables such as colliders, post-treatment variables, or non-confounders, 
such variables should of course be excluded from any propensity model. 

Researchers who are concerned about an underlying causal graph invalidating their propensity model 
can also turn to the causal discovery literature (i.e. \cite{peters2017elements}). 
Specifically, \cite{pmlr-v31-entner13a} present an approach to the problem of identifying confounders 
for adjustment in a causal model. In practice, the challenge of specifying (or estimating from the data) a 
causal graph can present a number of pitfalls beyond the scope of this paper. We suffice to say that 
there is no easy replacement for domain expertise in this process, but that this point applies to 
propensity models that use fully labelled data as well that those that use unlabeled data. 

In order to better understand the pitfalls of including non-confounders in a propensity model, 
we test several possibilities in a simulation study. Letting $\rvX_j$ be a variable in a set of $p$ covariates, 
we consider four cases:
\begin{enumerate}[a)]
\item \textbf{Case 1}: $\rvX_j$ is a confounder and is included in a propensity model
\item \textbf{Case 2}: $\rvX_j$ is \underline{\textbf{not}} a confounder and is included in a propensity model  
\item \textbf{Case 3}: $\rvX_j$ is a confounder and is \underline{\textbf{not}} included in a propensity model
\item \textbf{Case 4}: $\rvX_j$ is \underline{\textbf{not}} a confounder and is \underline{\textbf{not}} 
included in a propensity model
\end{enumerate}
Cases 1 and 4 are ideal, while cases 2 and 3 both fail to properly account for the true underlying 
causal model. The question we are concerned with is whether erring on the side of including 
non-confounders in a propensity model (case 2) is as harmful to inference as ignoring true 
confounders (case 3). To do this, we simulate 100 datasets of $n = $1,000 from the following 
data generating process:
\begin{align*}
\rvY &= \rvX\beta + \tau \rvZ + \varepsilon\\
\rvX, \varepsilon &\sim \mathcal{N}(0, 1)\\
\tau &= 3\\
p & = \left(\frac{1}{1+\exp(-\rvX\gamma)}\right)\\
\rvZ & \sim \textrm{Bernoulli}(p)
\end{align*}
For all variables except $\rvX_j$, the $\beta$ and $\gamma$ coefficients for the outcome and propensity 
models are drawn uniformly at random from a range of $(-0.5, 0.5)$ and $(-0.3, 0.3)$, respectively. 
For the scenarios in which $\rvX_j$ is a confounder (that is, $\rvX_j$ impacts both $\rvY$ and $\rvZ$), 
we set $\beta_{\rvX_j} = 0.5$. Each of our simulations assume some relationship between $\obsXvec_j$ and $\obsZ$, 
and we vary $\gamma_{\rvX_j}$ between $\{0, 1, 10\}$ to test the extent to which conditioning on $\rvX_j$ in a 
propensity model separates the treatment and control groups.

Below we see the simulated bias, RMSE, and 95\% interval coverage of each of the four cases, 
using BCF with a BART propensity model and $\gamma_{\rvX_j} = 1$
\begin{table}[ht]
\centering
\begingroup\small
\begin{tabular}{cccccc}
  \hline
Case & $\rvX_j$ confounder? & $\rvX_j$ in $\hat{p}$? & Bias & RMSE & Coverage \\ 
  \hline
1 & Yes & Yes & 0.004 & 0.098 & 89\% \\ 
2 & No & Yes & -0.008 & 0.096 & 88\% \\ 
3 & Yes & No & 0.068 & 0.109 & 84\% \\ 
4 & No & No & -0.007 & 0.087 & 87\% \\ 
   \hline
\end{tabular}
\endgroup
\caption{ATE estimation using BCF, $\gamma_{\rvX_j} = 1$} 
\end{table}

We see little difference in the results across each of the cases, because the influence of $\rvX_j$ on $\rvY$ is 
modest in the confounded case. Below is the same set of outputs for $\gamma_{\rvX_j} = 10$.

\begin{table}[ht]
\centering
\begingroup\small
\begin{tabular}{cccccc}
  \hline
Case & $\rvX_j$ confounder? & $\rvX_j$ in $\hat{p}$? & Bias & RMSE & Coverage \\ 
  \hline
1 & Yes & Yes & 0.033 & 0.253 & 91\% \\ 
2 & No & Yes & -0.09 & 0.262 & 87\% \\ 
3 & Yes & No & 0.24 & 0.252 & 80\% \\ 
4 & No & No & -0.045 & 0.126 & 98\% \\ 
   \hline
\end{tabular}
\endgroup
\caption{ATE estimation using BCF, $\gamma_{\rvX_j} = 10$} 
\end{table}

We see that cases 1 and 4 exhibit the best performance (since they properly incorporate the true 
causal graph), but even when $\rvX_j$ and $\rvZ$ are strongly related, including $\rvX_j$ in a propensity 
model when $\rvX_j$ is not a confounder is less harmful than excluding $\rvX_j$ when it is a confounder. 
To understand why, we briefly review the discussion of ``regularization-induced confounding" (RIC) 
covered in \cite{hahn2020bayesian}.

As we have seen in our simulations, classic parametric models of a treatment assignment are vulnerable to 
misspecification. While this shortcoming suggests the use of flexible machine learning methods to estimate 
propensity scores, such methods run the risk of overfitting the data and estimating propensities close to 0 or 1. 
In addition to numeric instability of weighting estimators, such overfitting can also violate the positivity or 
conditional exchangeability assumption. Researchers who want to fit flexible propensity models while avoiding 
overfitting can use regularized machine learning methods, such as BART. \cite{hahn2020bayesian}, building on 
\cite{hahn2018regularization}, show that a naive regularized model can bias treatment effect estimates. 
Their proposed solution, BCF, controls this bias by incorporating the estimated propensities as a feature in 
a regression model predicting $\E(\rvY \mid \rvZ = 0, \hat{p}(\rvX))$.

In our case, since the simulation studies above are conducted using BCF with a BART propensity model, 
we avoid some of the common problems of overfit propensity models while also 
recovering unbiased estimates of the average treatment effect. 
This explains why including a non-confounder, $\rvX_j$, which is strongly predictive of $\rvZ$ in the propensity 
model doesn't harm inference of the ATE by nearly as much as excluding $\rvX_j$ when it is a confounder. 

\section*{Acknowledgements}

This work was partially supported by NSF Grant DMS-1502640.

\bibliographystyle{imsart-nameyear} 
\bibliography{semi-supervised-propensity}

\begin{thebibliography}{25}
% BibTex style file: imsart-nameyear.bst, 2017-11-03
% Default style options (sort=1,type=nameyear).
% Used options (sort=1,type=nameyear).

\bibitem[\protect\citeauthoryear{Belkin, Niyogi and
  Sindhwani}{2006}]{belkin2006manifold}
\begin{barticle}[author]
\bauthor{\bsnm{Belkin},~\bfnm{Mikhail}\binits{M.}},
  \bauthor{\bsnm{Niyogi},~\bfnm{Partha}\binits{P.}} \AND
  \bauthor{\bsnm{Sindhwani},~\bfnm{Vikas}\binits{V.}}
(\byear{2006}).
\btitle{Manifold regularization: A geometric framework for learning from
  labeled and unlabeled examples}.
\bjournal{Journal of machine learning research}
\bvolume{7}
\bpages{2399--2434}.
\end{barticle}
\endbibitem

\bibitem[\protect\citeauthoryear{Boyd and Vandenberghe}{2004}]{boyd2004convex}
\begin{bbook}[author]
\bauthor{\bsnm{Boyd},~\bfnm{Stephen}\binits{S.}} \AND
  \bauthor{\bsnm{Vandenberghe},~\bfnm{Lieven}\binits{L.}}
(\byear{2004}).
\btitle{Convex optimization}.
\bpublisher{Cambridge university press}.
\end{bbook}
\endbibitem

\bibitem[\protect\citeauthoryear{Breiman}{2001}]{breiman2001random}
\begin{barticle}[author]
\bauthor{\bsnm{Breiman},~\bfnm{Leo}\binits{L.}}
(\byear{2001}).
\btitle{Random Forests}.
\bjournal{Machine Learning}
\bvolume{45}
\bpages{5--32}.
\end{barticle}
\endbibitem

\bibitem[\protect\citeauthoryear{Breiman
  et~al.}{1984}]{breiman1984classification}
\begin{barticle}[author]
\bauthor{\bsnm{Breiman},~\bfnm{Leo}\binits{L.}},
  \bauthor{\bsnm{Friedman},~\bfnm{Jerome~H}\binits{J.~H.}},
  \bauthor{\bsnm{Olshen},~\bfnm{Richard~A}\binits{R.~A.}} \AND
  \bauthor{\bsnm{Stone},~\bfnm{Charles~J}\binits{C.~J.}}
(\byear{1984}).
\btitle{Classification and Regression Trees}.
\end{barticle}
\endbibitem

\bibitem[\protect\citeauthoryear{Cerulli}{2014}]{cerulli2014treatrew}
\begin{barticle}[author]
\bauthor{\bsnm{Cerulli},~\bfnm{Giovanni}\binits{G.}}
(\byear{2014}).
\btitle{treatrew: A user-written command for estimating average treatment
  effects by reweighting on the propensity score}.
\bjournal{The Stata Journal}
\bvolume{14}
\bpages{541--561}.
\end{barticle}
\endbibitem

\bibitem[\protect\citeauthoryear{Chipman et~al.}{2010}]{chipman2010bart}
\begin{barticle}[author]
\bauthor{\bsnm{Chipman},~\bfnm{Hugh~A}\binits{H.~A.}},
  \bauthor{\bsnm{George},~\bfnm{Edward~I}\binits{E.~I.}},
  \bauthor{\bsnm{McCulloch},~\bfnm{Robert~E}\binits{R.~E.}} \betal{et~al.}
(\byear{2010}).
\btitle{BART: Bayesian additive regression trees}.
\bjournal{The Annals of Applied Statistics}
\bvolume{4}
\bpages{266--298}.
\end{barticle}
\endbibitem

\bibitem[\protect\citeauthoryear{Entner, Hoyer and
  Spirtes}{2013}]{pmlr-v31-entner13a}
\begin{binproceedings}[author]
\bauthor{\bsnm{Entner},~\bfnm{Doris}\binits{D.}},
  \bauthor{\bsnm{Hoyer},~\bfnm{Patrik}\binits{P.}} \AND
  \bauthor{\bsnm{Spirtes},~\bfnm{Peter}\binits{P.}}
(\byear{2013}).
\btitle{Data-driven covariate selection for nonparametric estimation of causal
  effects}.
In \bbooktitle{Proceedings of the Sixteenth International Conference on
  Artificial Intelligence and Statistics}
(\beditor{\bfnm{Carlos~M.}\binits{C.~M.}~\bsnm{Carvalho}} \AND
  \beditor{\bfnm{Pradeep}\binits{P.}~\bsnm{Ravikumar}}, eds.).
\bseries{Proceedings of Machine Learning Research}
\bvolume{31}
\bpages{256--264}.
\bpublisher{PMLR}.
\end{binproceedings}
\endbibitem

\bibitem[\protect\citeauthoryear{Gruber and van~der
  Laan}{2009}]{gruber2009targeted}
\begin{barticle}[author]
\bauthor{\bsnm{Gruber},~\bfnm{Susan}\binits{S.}} \AND
  \bauthor{\bparticle{van~der} \bsnm{Laan},~\bfnm{Mark~J}\binits{M.~J.}}
(\byear{2009}).
\btitle{Targeted maximum likelihood estimation: A gentle introduction}.
\end{barticle}
\endbibitem

\bibitem[\protect\citeauthoryear{Hahn et~al.}{2018}]{hahn2018regularization}
\begin{barticle}[author]
\bauthor{\bsnm{Hahn},~\bfnm{P~Richard}\binits{P.~R.}},
  \bauthor{\bsnm{Carvalho},~\bfnm{Carlos~M}\binits{C.~M.}},
  \bauthor{\bsnm{Puelz},~\bfnm{David}\binits{D.}},
  \bauthor{\bsnm{He},~\bfnm{Jingyu}\binits{J.}} \betal{et~al.}
(\byear{2018}).
\btitle{Regularization and confounding in linear regression for treatment
  effect estimation}.
\bjournal{Bayesian Analysis}
\bvolume{13}
\bpages{163--182}.
\end{barticle}
\endbibitem

\bibitem[\protect\citeauthoryear{Hahn et~al.}{2020}]{hahn2020bayesian}
\begin{barticle}[author]
\bauthor{\bsnm{Hahn},~\bfnm{P~Richard}\binits{P.~R.}},
  \bauthor{\bsnm{Murray},~\bfnm{Jared~S}\binits{J.~S.}},
  \bauthor{\bsnm{Carvalho},~\bfnm{Carlos~M}\binits{C.~M.}} \betal{et~al.}
(\byear{2020}).
\btitle{Bayesian regression tree models for causal inference: regularization,
  confounding, and heterogeneous effects}.
\bjournal{Bayesian Analysis}.
\end{barticle}
\endbibitem

\bibitem[\protect\citeauthoryear{Harville}{1998}]{harville1998matrix}
\begin{bmisc}[author]
\bauthor{\bsnm{Harville},~\bfnm{David~A}\binits{D.~A.}}
(\byear{1998}).
\btitle{Matrix algebra from a statistician's perspective}.
\end{bmisc}
\endbibitem

\bibitem[\protect\citeauthoryear{Hernan and Robins}{2020}]{hernan2020causal}
\begin{bbook}[author]
\bauthor{\bsnm{Hernan},~\bfnm{Miguel~A}\binits{M.~A.}} \AND
  \bauthor{\bsnm{Robins},~\bfnm{James~M}\binits{J.~M.}}
(\byear{2020}).
\btitle{Causal Inference: What If}.
\bpublisher{Boca Raton: Chapman \& Hall, CRC}.
\end{bbook}
\endbibitem

\bibitem[\protect\citeauthoryear{Hirano, Imbens and
  Ridder}{2003}]{hirano2003efficient}
\begin{barticle}[author]
\bauthor{\bsnm{Hirano},~\bfnm{Keisuke}\binits{K.}},
  \bauthor{\bsnm{Imbens},~\bfnm{Guido~W}\binits{G.~W.}} \AND
  \bauthor{\bsnm{Ridder},~\bfnm{Geert}\binits{G.}}
(\byear{2003}).
\btitle{Efficient estimation of average treatment effects using the estimated
  propensity score}.
\bjournal{Econometrica}
\bvolume{71}
\bpages{1161--1189}.
\end{barticle}
\endbibitem

\bibitem[\protect\citeauthoryear{Lee, Lessler and
  Stuart}{2010}]{lee2010improving}
\begin{barticle}[author]
\bauthor{\bsnm{Lee},~\bfnm{Brian~K}\binits{B.~K.}},
  \bauthor{\bsnm{Lessler},~\bfnm{Justin}\binits{J.}} \AND
  \bauthor{\bsnm{Stuart},~\bfnm{Elizabeth~A}\binits{E.~A.}}
(\byear{2010}).
\btitle{Improving propensity score weighting using machine learning}.
\bjournal{Statistics in medicine}
\bvolume{29}
\bpages{337--346}.
\end{barticle}
\endbibitem

\bibitem[\protect\citeauthoryear{Liang, Mukherjee and
  West}{2007}]{liang2007use}
\begin{barticle}[author]
\bauthor{\bsnm{Liang},~\bfnm{Feng}\binits{F.}},
  \bauthor{\bsnm{Mukherjee},~\bfnm{Sayan}\binits{S.}} \AND
  \bauthor{\bsnm{West},~\bfnm{Mike}\binits{M.}}
(\byear{2007}).
\btitle{The use of unlabeled data in predictive modeling}.
\bjournal{Statistical Science}
\bpages{189--205}.
\end{barticle}
\endbibitem

\bibitem[\protect\citeauthoryear{Little and
  Rubin}{2002}]{little2019statistical}
\begin{bbook}[author]
\bauthor{\bsnm{Little},~\bfnm{Roderick~JA}\binits{R.~J.}} \AND
  \bauthor{\bsnm{Rubin},~\bfnm{Donald~B}\binits{D.~B.}}
(\byear{2002}).
\btitle{Statistical analysis with missing data}.
\bpublisher{John Wiley \& Sons}.
\end{bbook}
\endbibitem

\bibitem[\protect\citeauthoryear{Lunceford and
  Davidian}{2004}]{lunceford2004stratification}
\begin{barticle}[author]
\bauthor{\bsnm{Lunceford},~\bfnm{Jared~K}\binits{J.~K.}} \AND
  \bauthor{\bsnm{Davidian},~\bfnm{Marie}\binits{M.}}
(\byear{2004}).
\btitle{Stratification and weighting via the propensity score in estimation of
  causal treatment effects: a comparative study}.
\bjournal{Statistics in medicine}
\bvolume{23}
\bpages{2937--2960}.
\end{barticle}
\endbibitem

\bibitem[\protect\citeauthoryear{McCaffrey, Ridgeway and
  Morral}{2004}]{mccaffrey2004propensity}
\begin{barticle}[author]
\bauthor{\bsnm{McCaffrey},~\bfnm{Daniel~F}\binits{D.~F.}},
  \bauthor{\bsnm{Ridgeway},~\bfnm{Greg}\binits{G.}} \AND
  \bauthor{\bsnm{Morral},~\bfnm{Andrew~R}\binits{A.~R.}}
(\byear{2004}).
\btitle{Propensity score estimation with boosted regression for evaluating
  causal effects in observational studies.}
\bjournal{Psychological methods}
\bvolume{9}
\bpages{403}.
\end{barticle}
\endbibitem

\bibitem[\protect\citeauthoryear{Peters, Janzing and
  Sch{\"o}lkopf}{2017}]{peters2017elements}
\begin{bbook}[author]
\bauthor{\bsnm{Peters},~\bfnm{Jonas}\binits{J.}},
  \bauthor{\bsnm{Janzing},~\bfnm{Dominik}\binits{D.}} \AND
  \bauthor{\bsnm{Sch{\"o}lkopf},~\bfnm{Bernhard}\binits{B.}}
(\byear{2017}).
\btitle{Elements of causal inference: foundations and learning algorithms}.
\bpublisher{MIT press}.
\end{bbook}
\endbibitem

\bibitem[\protect\citeauthoryear{Rosenbaum}{1987}]{rosenbaum1987model}
\begin{barticle}[author]
\bauthor{\bsnm{Rosenbaum},~\bfnm{Paul~R}\binits{P.~R.}}
(\byear{1987}).
\btitle{Model-based direct adjustment}.
\bjournal{Journal of the American Statistical Association}
\bvolume{82}
\bpages{387--394}.
\end{barticle}
\endbibitem

\bibitem[\protect\citeauthoryear{Rosenbaum and
  Rubin}{1983}]{rosenbaum1983central}
\begin{barticle}[author]
\bauthor{\bsnm{Rosenbaum},~\bfnm{Paul~R}\binits{P.~R.}} \AND
  \bauthor{\bsnm{Rubin},~\bfnm{Donald~B}\binits{D.~B.}}
(\byear{1983}).
\btitle{The central role of the propensity score in observational studies for
  causal effects}.
\bjournal{Biometrika}
\bvolume{70}
\bpages{41--55}.
\end{barticle}
\endbibitem

\bibitem[\protect\citeauthoryear{Rubin}{1980}]{rubin1980randomization}
\begin{barticle}[author]
\bauthor{\bsnm{Rubin},~\bfnm{Donald~B}\binits{D.~B.}}
(\byear{1980}).
\btitle{Randomization analysis of experimental data: The Fisher randomization
  test comment}.
\bjournal{Journal of the American Statistical Association}
\bvolume{75}
\bpages{591--593}.
\end{barticle}
\endbibitem

\bibitem[\protect\citeauthoryear{van~der Laan}{2010a}]{van2010targeted}
\begin{barticle}[author]
\bauthor{\bparticle{van~der} \bsnm{Laan},~\bfnm{Mark~J}\binits{M.~J.}}
(\byear{2010}a).
\btitle{Targeted Maximum Likelihood Based Causal Inference: Part I}.
\bjournal{The International Journal of Biostatistics}
\bvolume{6}.
\end{barticle}
\endbibitem

\bibitem[\protect\citeauthoryear{van~der Laan}{2010b}]{van2010targeted2}
\begin{barticle}[author]
\bauthor{\bparticle{van~der} \bsnm{Laan},~\bfnm{Mark~J}\binits{M.~J.}}
(\byear{2010}b).
\btitle{Targeted Maximum Likelihood Based Causal Inference: Part II}.
\bjournal{The International Journal of Biostatistics}
\bvolume{6}.
\end{barticle}
\endbibitem

\bibitem[\protect\citeauthoryear{Zhu and Goldberg}{2009}]{zhu2009introduction}
\begin{barticle}[author]
\bauthor{\bsnm{Zhu},~\bfnm{Xiaojin}\binits{X.}} \AND
  \bauthor{\bsnm{Goldberg},~\bfnm{Andrew~B}\binits{A.~B.}}
(\byear{2009}).
\btitle{Introduction to semi-supervised learning}.
\bjournal{Synthesis lectures on artificial intelligence and machine learning}
\bvolume{3}
\bpages{1--130}.
\end{barticle}
\endbibitem

\end{thebibliography}

\begin{appendix}
\section{Variance derivations for $\mbox{IPW}_{p(X)}$ vs $\mbox{IPW}_{\hat{p}(x)}$}\label{appA}

Consider the data generating process of Section \ref{trueprop}. We condition on a sample of size $\obsN$ 
and observed values of $\obsXmat$, such that $\obsN_x$ is also observed. We derive the variance of the 
IPW estimators for an arbitrary $\{X = x\}$ stratum and then conclude that the result holds across the support of $X$.
We randomize over the sample sizes of treated units, which we model as $N_{x, z=1} \sim \mbox{Binomial}(\obsN_x, p_x)$:

\begin{equation*}
\begin{aligned}
\mbox{V}\left[\mbox{IPW}_{p(x), x}\right] &= \mbox{V} \left[ \left( \frac{\obsN_{x}}{\obsN} \right) \left( \frac{\hat{p}_{x}}{p_{x}} \bar{Y}_{x, z=1} - \frac{1-\hat{p}_{x}}{1-p_{x}} \bar{Y}_{x, z=0} \right) \right]\\ 
&= \left(\frac{\obsN_{x}}{\obsN}\right)^2 \mbox{V} \left[ \frac{ \hat{p}_{x}}{p_{x}} \bar{Y}_{x, z=1} - \frac{ 1 - \hat{p}_{x}}{1 - p_{x}} \bar{Y}_{x, z=0} \right] \\
&= \left(\frac{\obsN_{x}}{\obsN}\right)^2 \mbox{V} \left[ \frac{ N_{x, z=1}}{ \obsN_{x} p_{x}} \bar{Y}_{x, z=1} - \frac{ \obsN_{x} - N_{x, z=1}}{\obsN_{x}(1 - p_{x})} \bar{Y}_{x, z=0} \right] \\
&= \left(\frac{\obsN_{x}}{\obsN}\right)^2 \mbox{V} \left[ \frac{ N_{x, z=1}}{ \obsN_{x} p_{x}} \bar{Y}_{x, z=1} - \frac{ \obsN_{x} - N_{x, z=1}}{\obsN_{x}(1 - p_{x})} \bar{Y}_{x, z=0} \right] \\
&= \left(\frac{\obsN_{x}}{\obsN}\right)^2 \mbox{E} \left[ \mbox{V} \left[ \frac{ N_{x, z=1}}{ \obsN_{x} p_{x}} \bar{Y}_{x, z=1} - \frac{ \obsN_{x} - N_{x, z=1}}{\obsN_{x}(1 - p_{x})} \bar{Y}_{x, z=0} \mid N_{x, z=1} \right] \right]\\ 
&\;\;\;\; + \left(\frac{\obsN_{x}}{\obsN}\right)^2 \mbox{V} \left[ \mbox{E} \left[ \frac{ N_{x, z=1}}{ \obsN_{x} p_{x}} \bar{Y}_{x, z=1} - \frac{ \obsN_{x} - N_{x, z=1}}{\obsN_{x}(1 - p_{x})} \bar{Y}_{x, z=0} \mid N_{x, z=1} \right] \right] \\
&= \left(\frac{\obsN_{x}}{\obsN}\right)^2 \mbox{E} \left[ \left(\frac{ N_{x, z=1}}{ \obsN_{x} p_{x}}\right)^2 \mbox{V} \left[ \bar{Y}_{x, z=1} \mid N_{x, z=1} \right] + \left(\frac{ \obsN_{x} - N_{x, z=1}}{\obsN_{x}(1 - p_{x})}\right)^2 \V \left [\bar{Y}_{x, z=0} \mid N_{x, z=1} \right] \right]\\ 
&\;\;\;\; + \left(\frac{\obsN_{x}}{\obsN}\right)^2 \mbox{V} \left[ \mbox{E} \left[ \frac{ N_{x, z=1}}{ \obsN_{x} p_{x}} \bar{Y}_{x, z=1} - \frac{ \obsN_{x} - N_{x, z=1}}{\obsN_{x}(1 - p_{x})} \bar{Y}_{x, z=0} \mid N_{x, z=1} \right] \right] \\
&= \left(\frac{\obsN_{x}}{\obsN}\right)^2 \left(\frac{ \obsN_{x} p_{x} (1 - p_{x}) + \obsN_{x}^2 p_{x}^2 }{ \obsN_{x}^2 p_{x}^2}\mbox{E} \left[  \V \left[ \bar{Y}_{x, z=1} \mid N_{x, z=1} \right] \right] \right)\\
&\;\;\;\; + \left(\frac{\obsN_{x}}{\obsN}\right)^2 \left(\frac{ \obsN_{x} p_{x} (1 - p_{x}) + \obsN_{x}^2 (1-p_{x})^2}{\obsN_{x}^2(1 - p_{x})^2} \mbox{E} \left[ \V \left [\bar{Y}_{x, z=0} \mid N_{x, z=1} \right]\right] \right)\\ 
&\;\;\;\; + \left(\frac{\obsN_{x}}{\obsN}\right)^2 \mbox{V} \left[ \mbox{E} \left[ \frac{ N_{x, z=1}}{ \obsN_{x} p_{x}} \bar{Y}_{x, z=1} - \frac{ \obsN_{x} - N_{x, z=1}}{\obsN_{x}(1 - p_{x})} \bar{Y}_{x, z=0} \mid N_{x, z=1} \right] \right] \\
\end{aligned}
\end{equation*}
while
\begin{equation*}
\begin{aligned}
\mbox{V}\left[\mbox{IPW}_{\hat{p}(x), x}\right] &= \V \left[ \left( \frac{\obsN_{x}}{\obsN} \right) \left( \bar{Y}_{x, z=1} 
- \bar{Y}_{x, z=0} \right) \right]\\ 
&= \left(\frac{\obsN_{x}}{\obsN}\right)^2 \V \left[ \bar{Y}_{x, z=1} - \bar{Y}_{x, z=0} \right] \\
&= \left(\frac{\obsN_{x}}{\obsN}\right)^2 \E \left[ \V \left[ \bar{Y}_{x, z=1} - \bar{Y}_{x, z=0} \mid N_{x, z=1} \right] \right]\\
&\;\;\;\; + \left(\frac{\obsN_{x}}{\obsN}\right)^2 \V \left[ \E \left[ \bar{Y}_{x, z=1} - \bar{Y}_{x, z=0} \mid N_{x, z=1} \right] \right]\\
&= \left(\frac{\obsN_{x}}{\obsN}\right)^2 \E \left[ \V \left[ \bar{Y}_{x, z=1} - \bar{Y}_{x, z=0} \mid N_{x, z=1} \right] \right]\\
&\;\;\;\; + \left(\frac{\obsN_{x}}{\obsN}\right)^2 \V \left[ \tau \right]\\
&= \left(\frac{\obsN_{x}}{\obsN}\right)^2 \E \left[ \V \left[ \bar{Y}_{x, z=1} - \bar{Y}_{x, z=0} \mid N_{x, z=1} \right] \right]\\
\end{aligned}
\end{equation*}

Since variances are nonnegative and both 
$\frac{ \obsN_{x} p_{x} (1 - p_{x}) + \obsN_{x}^2 p_{x}^2 }{ \obsN_{x}^2 p_{x}^2} > 1$ and 
$\frac{ \obsN_{x} p_{x} (1 - p_{x}) + \obsN_{x}^2 (1-p_{x})^2}{\obsN_{x}^2(1 - p_{x})^2} > 1$ for 
$\obsN_{x} > 0$ and $0 < p_{x} < 1$, we see that 
$\mbox{V}\left[\mbox{IPW}_{p(x), x}\right] > \mbox{V}\left[\mbox{IPW}_{\hat{p}(x), x}\right]$. 
Since $X = x$ has been arbitrary, this result applies to all $\{X = x\}$ strata and thus it holds 
in expectation across the support of $X$.

\section{Variance derivations for $\mbox{IPW}_{\hat{p}(x)}$ vs $\mbox{IPW}_{\hat{p}(p(x))}$}\label{appB}

Now, we compare the variance of the $\hat{p}(p(X))$ estimator with that of the $\hat{p}(X)$ estimator. 
These two estimators differ only in their use of $X_2$. 
Since $p(X) = f(X_1)$, the $\hat{p}(p(X))$ estimator is equivalent to a stratification estimator using only 
$X_1$, while the $\hat{p}(X)$ estimator stratifies on both $X_1$ and $X_2$.
In this derivation, we condition on a sample of size $\obsN$ and observed values of $\obsXvec_1$ and $\obsZ$, so that 
$\obsN_{x_1, z}$ is observed. We randomize over $X_2$, deriving the variance of the IPW estimators for an 
arbitrary $\{\rvX_1 = x_1, \rvZ = 1\}$ stratum and concluding that the result holds across the support of $X$ and $Z$.
To represent the randomization over $X_2$ at the level of a $\{X_1 = x_1\}$ stratum, we introduce two binomial random variables:
\begin{itemize}
\item $\rvN_{x_1, \rvX_2 = 1, \obsZ=1} \sim \mbox{Binomial}(\obsN_{x_1, \obsZ=1}, 1/2)$
\item $\rvN_{x_1, \rvX_2 = 1, \obsZ=0} \sim \mbox{Binomial}(\obsN_{x_1, \obsZ=0}, 1/2)$
\end{itemize}

We can decompose a given $\{X_1 = x_1\}$ stratum of the $\hat{p}(p(X))$ estimator as follows
\begin{equation*}
\begin{aligned}
\mbox{IPW}_{\hat{p}(p(x)), x_1} &= \frac{1}{\obsN} \sum_{j: X_{1,j} = x_1} \left(  \frac{Y_i Z_i}{\hat{p}(p_{x_1})} - 
\frac{Y_i (1-Z_i)}{1-\hat{p}(p_{x_1})} \right)\\ 
&= \frac{\obsN_{x_1}}{\obsN} \left( \frac{\rvN_{x_1, \rvX_2 = 1, \obsZ=1}}{\obsN_{x_1, z=1}} \bar{Y}_{X_1 = x_1, X_2 = 1, Z=1} + 
\frac{\obsN_{x_1, z=1} - \rvN_{x_1, \rvX_2 = 1, \obsZ=1}}{\obsN_{x_1, z=1}} \bar{Y}_{X_1 = x_1, X_2 = 0, Z=1} \right)\\
&\;\;\;\; - \frac{\obsN_{x_1}}{\obsN} \left( \frac{\rvN_{x_1, \rvX_2 = 1, \obsZ=0}}{\obsN_{x_1, z=0}} \bar{Y}_{X_1 = x_1, X_2 = 1, Z=0} 
+ \frac{\obsN_{x_1, z=0} - \rvN_{x_1, \rvX_2 = 1, \obsZ=0}}{\obsN_{x_1, z=0}} \bar{Y}_{X_1 = x_1, X_2 = 0, Z=0} \right)\\
\end{aligned}
\end{equation*}

We compute the variance of this stratum-specific estimator via the law of total variance
\begin{equation*}
\begin{aligned}
\V \left[ \mbox{IPW}_{\hat{p}(p(x)), x_1} \right] &= \E \left[ \V \left[ \mbox{IPW}_{\hat{p}(p(x)), x_1} \mid 
\rvN_{x_1, x_2 = 1, z=1}, \rvN_{x_1, x_2 = 1, z=0} \right] \right]\\
&\;\;\;\; + \V \left[ \E \left[ \mbox{IPW}_{\hat{p}(p(x)), x_1} \mid 
\rvN_{x_1, x_2 = 1, z=1}, \rvN_{x_1, x_2 = 1, z=0} \right] \right]\\ 
\end{aligned}
\end{equation*}
where
\begin{equation*}
\begin{aligned}
&\E \left[ \mbox{IPW}_{\hat{p}(p(x)), x_1} \mid \rvN_{x_1, \rvX_2 = 1, \obsZ=1}, \rvN_{x_1, \rvX_2 = 1, \obsZ=0} \right] = \\
&\;\;\;\;\frac{\obsN_{x_1}}{\obsN} \left( \frac{\rvN_{x_1, \rvX_2 = 1, \obsZ=1}}{\obsN_{x_1, z=1}} 
\E \left[\bar{Y}_{X_1 = x_1, X_2 = 1, Z=1}\right] + \frac{\obsN_{x_1, z=1} - \rvN_{x_1, \rvX_2 = 1, \obsZ=1}}{\obsN_{x_1, z=1}} 
\E \left[ \bar{Y}_{X_1 = x_1, X_2 = 0, Z=1} \right] \right)\\
&\;\;\;\;\;\;\;\; - \frac{\obsN_{x_1}}{\obsN} \left( \frac{\rvN_{x_1, \rvX_2 = 1, \obsZ=0}}{\obsN_{x_1, z=0}} 
\E \left[ \bar{Y}_{X_1 = x_1, X_2 = 1, Z=0} \right] + \frac{\obsN_{x_1, z=0} - \rvN_{x_1, \rvX_2 = 1, \obsZ=0}}{\obsN_{x_1, z=0}} 
\E \left[ \bar{Y}_{X_1 = x_1, X_2 = 0, Z=0} \right] \right)\\
\end{aligned}
\end{equation*}
If $X_2$ has no pure prognostic / moderating impact on $Y$, then it is true that 
\begin{itemize}
\item $\E \left[\bar{Y}_{X_1 = x_1, X_2 = 1, Z=1}\right] = \E \left[\bar{Y}_{X_1 = x_1, X_2 = 0, Z=1}\right]$ and 
\item $\E \left[\bar{Y}_{X_1 = x_1, X_2 = 1, Z=0}\right] = \E \left[\bar{Y}_{X_1 = x_1, X_2 = 0, Z=0}\right]$,
\end{itemize}
and thus $\V \left[ \E \left[ \mbox{IPW}_{\hat{p}(p(x)), x_1} \mid \rvN_{x_1, \rvX_2 = 1, \obsZ=1}, \rvN_{x_1, \rvX_2 = 1, \obsZ=0} \right]\right] = 0$.

So the variance of $\mbox{IPW}_{\hat{p}(p(x)), x_1}$ is equal to $\E\left[ \V \left[ \mbox{IPW}_{\hat{p}(p(x)), x_1} \mid \rvN_{x_1, \rvX_2 = 1, \obsZ=1}, \rvN_{x_1, \rvX_2 = 1, \obsZ=0} \right] \right]$ when $X_2$ is not a pure prognostic / moderator variable.

Now, observe that a given $\{X_1 = x_1\}$ stratum of the $\hat{p}(X)$ estimator can be written as
\begin{equation*}
\begin{aligned}
\mbox{IPW}_{\hat{p}(x), x_1} &= \frac{1}{\obsN} \sum_{j: X_{1,j} = x_1} \left(  \frac{Y_i Z_i}{\hat{p}(X_i)} - \frac{Y_i (1-Z_i)}{1-\hat{p}(X_i)} \right)\\ 
&= \frac{N_{x_1, \rvX_2=1}}{\obsN} \bar{Y}_{X_1 = x_1, X_2 = 1, Z=1} + \frac{N_{x_1, \rvX_2=0}}{\obsN} \bar{Y}_{X_1 = x_1, X_2 = 0, Z=1}\\
&\;\;\;\; - \frac{N_{x_1, \rvX_2=1}}{\obsN} \bar{Y}_{X_1 = x_1, X_2 = 1, Z=0} - \frac{N_{x_1, \rvX_2=0}}{\obsN} \bar{Y}_{X_1 = x_1, X_2 = 0, Z=0}\\
\end{aligned}
\end{equation*}

To obtain the variance of this estimator, we again apply the law of total variance
\begin{equation*}
\begin{aligned}
\V \left[ \mbox{IPW}_{\hat{p}(x), x_1} \right] &= \E \left[ \V \left[ \mbox{IPW}_{\hat{p}(x), x_1} \mid \rvN_{x_1, \rvX_2 = 1, \obsZ=1}, \rvN_{x_1, \rvX_2 = 1, \obsZ=0} \right] \right]\\
&\;\;\;\; + \V \left[ \E \left[ \mbox{IPW}_{\hat{p}(x), x_1} \mid \rvN_{x_1, \rvX_2 = 1, \obsZ=1}, \rvN_{x_1, \rvX_2 = 1, \obsZ=0} \right] \right]\\ 
\end{aligned}
\end{equation*}

and
\begin{equation*}
\begin{aligned}
&\E \left[ \mbox{IPW}_{\hat{p}(x), x_1} \mid \rvN_{x_1, \rvX_2 = 1, \obsZ=1}, \rvN_{x_1, \rvX_2 = 1, \obsZ=0} \right] = \\
&= \frac{N_{x_1, \rvX_2=1}}{\obsN} \E \left[ \bar{Y}_{X_1 = x_1, X_2 = 1, Z=1} \right] + \frac{\rvN_{x_1, x_2=0}}{\obsN} \E \left[\bar{Y}_{X_1 = x_1, X_2 = 0, Z=1}\right]\\
&\;\;\;\; - \frac{\rvN_{x_1, \rvX_2=1}}{\obsN} \E \left[ \bar{Y}_{X_1 = x_1, X_2 = 1, Z=0} \right] - \frac{\rvN_{x_1, \rvX_2=0}}{\obsN}\E \left[ \bar{Y}_{X_1 = x_1, X_2 = 0, Z=0} \right]\\
&= \frac{\rvN_{x_1, \rvX_2=1, \obsZ=0}+\rvN_{x_1, \rvX_2=1, \obsZ=1}}{\obsN} \left( \E \left[ \bar{Y}_{X_1 = x_1, X_2 = 1, Z=1} \right] - \E \left[ \bar{Y}_{X_1 = x_1, X_2 = 1, Z=0} \right] \right)\\
&\;\;\;\; + \frac{\rvN_{x_1, \rvX_2=0, \obsZ=0}+\rvN_{x_1, \rvX_2=0, \obsZ=1}}{\obsN} \left( \E \left[ \bar{Y}_{X_1 = x_1, X_2 = 0, Z=1} \right] - \E \left[ \bar{Y}_{X_1 = x_1, X_2 = 0, Z=0} \right] \right)
\end{aligned}
\end{equation*}
$\E \left[ \bar{Y}_{x_1, x_2, Z=1} \right] - \E \left[ \bar{Y}_{x_1, x_2, Z=0} \right] = \tau$ for $x_2 \in \{ 0, 1 \}$, so $\E \left[ \mbox{IPW}_{\hat{p}(x), x_1} \mid \rvN_{x_1, \rvX_2 = 1, \obsZ=1}, \rvN_{x_1, \rvX_2 = 1, \obsZ=0} \right] = \tau \obsN_{x_1} / \obsN$ and thus $\V \left[ \E \left[ \mbox{IPW}_{\hat{p}(x), x_1} \mid \rvN_{x_1, x_2 = 1, z=1}, \rvN_{x_1, x_2 = 1, z=0} \right] \right] = 0$ regardless of whether or not $X_2$ is a pure prognostic / moderator variable.

We can compare the variances of $\hat{p}(p(x))$ and $\hat{p}(x)$ without evaluating the expectation over $X_2$ for $\hat{p}(x)$ by comparing the conditional variance terms of the two expressions.

Focusing on the first two terms of the conditional variance expressions (corresponding to $X_1 = x_1$ and $Z=1$), we see that 
\begin{equation*}
\begin{aligned}
&\V \left[ \mbox{IPW}_{\hat{p}(x), x_1, z=1} \mid \rvN_{x_1, \rvX_2 = 1, \obsZ=1} \right]\\
&= \left(\frac{\rvN_{x_1, \rvX_2=1}}{\obsN} \right)^2 \frac{\V \left[ Y_{X_1 = x_1, X_2 = 1, Z=1} \right]}{\rvN_{x_1, \rvX_2 = 1, \obsZ=1}} + \left(\frac{\rvN_{x_1, \rvX_2=0}}{\obsN} \right)^2 \frac{ \V \left[ Y_{\rvX_1 = x_1, \rvX_2 = 0, Z=1} \right] }{\rvN_{x_1, \rvX_2 = 0, \obsZ=1}}\\
&= \frac{1}{\obsN^2} \frac{\left(\rvN_{x_1, \rvX_2=1, \obsZ=1} + \rvN_{x_1, \rvX_2=1, \obsZ=0} \right)^2 \V \left[ Y_{\rvX_1 = x_1, \rvX_2 = 1, \rvZ=1} \right]}{\rvN_{x_1, \rvX_2 = 1, \obsZ=1}}\\ 
&\;\;\;\;\;\;+ \frac{1}{\obsN^2} \frac{\left(\rvN_{x_1, \rvX_2=0, \obsZ=1} + \rvN_{x_1, \rvX_2=0, \obsZ=0} \right)^2 \V \left[ Y_{\rvX_1 = x_1, \rvX_2 = 0, \rvZ=1} \right]}{\rvN_{x_1, \rvX_2 = 0, \obsZ=1}}\\
\end{aligned}
\end{equation*}
and also that
\begin{equation*}
\begin{aligned}
&\V \left[ \mbox{IPW}_{\hat{p}(p(x)), x_1, z=1} \mid \rvN_{x_1, \rvX_2 = 1, \obsZ=1} \right]\\
&= \frac{1}{\obsN^2} \left[ \frac{\left( \obsN_{x_1} \right)^2 \left( \rvN_{x_1, \rvX_2=1, \obsZ=1} \V \left[ Y_{\rvX_1 = x_1, \rvX_2 = 1, \rvZ=1} \right] + \rvN_{x_1, \rvX_2=0, \obsZ=1} \V \left[ Y_{\rvX_1 = x_1, \rvX_2 = 0, \rvZ=1} \right] \right) }{ \left( \rvN_{x_1, \rvX_2 = 1, \obsZ=1} + \rvN_{x_1, \rvX_2 = 0, \obsZ=1} \right)^2}\right]\\
\end{aligned}
\end{equation*}
where $\obsN_{x_1} = \left( N_{x_1, \rvX_2=1, \obsZ=1} + N_{x_1, \rvX_2=1, \obsZ=0} \right) + \left( N_{x_1, \rvX_2=0, \obsZ=1} + N_{x_1, \rvX_2=0, \obsZ=0} \right)$

If $\V \left[ Y_{X_1 = x_1, X_2 = 1, Z=1} \right] = \V \left[ Y_{X_1 = x_1, X_2 = 0, Z=1} \right]$ then we can set $v = \V \left[ Y_{X_1 = x_1, X_2 = 1, Z=1} \right] = \V \left[ Y_{X_1 = x_1, X_2 = 0, Z=1} \right]$ and factor this term out.
After this adjustment, the comparison reduces to evaluating the conditions in which 
\begin{equation*}
\begin{aligned}
&\frac{ \obsN_{x_1} ^2}{N_{x_1, \rvX_2 = 1, \obsZ=1}+N_{x_1, \rvX_2 = 0, \obsZ=1}} \leq \frac{N_{x_1, \rvX_2=1}^2}{N_{x_1, \rvX_2 = 1, \obsZ=1}} + \frac{N_{x_1, \rvX_2=0}^2 }{N_{x_1, \rvX_2 = 0, \obsZ=1}}
\end{aligned}
\end{equation*}
where $N_{x_1, \rvX_2=1} = \left(N_{x_1, \rvX_2=1, \obsZ=1} + N_{x_1, \rvX_2=1, \obsZ=0}\right)$ and $N_{x_1, \rvX_2=0} = \left(N_{x_1, \rvX_2=0, \obsZ=1} + N_{x_1, \rvX_2=0, \obsZ=0}\right)$

Multiplying through by $1/2$, we see that this is true if the function $f(x, y) = \frac{(x+y)^2}{y}$ is convex. 
Let the domain equal $\R^2_+$ and observe that a sufficient condition for the convexity of $f$ is the 
positive definiteness of the Hessian matrix (\cite{boyd2004convex}).
\begin{equation*}
\begin{aligned}
H &= 
\begin{pmatrix}
\frac{\partial^2 f}{\partial x^2} & \frac{\partial^2 f}{\partial y\partial x} \\
\frac{\partial^2 f}{\partial y\partial x} & \frac{\partial^2 f}{\partial y^2}
\end{pmatrix} = 
\begin{pmatrix}
\frac{2}{y} & \frac{-2x}{y^2} \\
\frac{-2x}{y^2} & \frac{2x^2}{y^3}
\end{pmatrix}
\end{aligned}
\end{equation*}

This can be confirmed by observing that the three principal minors are nonnegative (\cite{harville1998matrix}):
\[\frac{2}{y} \geq 0; \;\;\;\; \frac{2x^2}{y^3} \geq 0; \;\;\;\; \frac{4x^2}{y^4} - \frac{4x^2}{y^4} = 0\]

Thus, it follows that 
\begin{equation*}
\begin{aligned}
&\frac{ \obsN_{x_1} ^2}{N_{x_1, \rvX_2 = 1, \obsZ=1}+N_{x_1, \rvX_2 = 0, \obsZ=1}} \leq \frac{N_{x_1, \rvX_2=1}^2}{N_{x_1, \rvX_2 = 1, \obsZ=1}} + \frac{N_{x_1, \rvX_2=0}^2 }{N_{x_1, \rvX_2 = 0, \obsZ=1}}
\end{aligned}
\end{equation*}
The positivity constraint on the convexity of $f$ translates into an assumption of ``no empty strata'' for our weighting estimator. 
Since this result holds for an arbitrary $\obsN_{x_1}$, it holds across the distribution of sample sizes for which no strata cells 
are empty.

\end{appendix}
\end{document}